# Towards Extended Active Learning for Modelling Ferroelectric Switching: the Need for "Gold Standards"

Jeffrey R. Reimers,*[ab] Wenbin Zhang,[a] Zhe Su,[a] Musen Li,[a,b,e] Carla Verdi,[c] Tim Gould,[d] and Wei Ren[a,e]

[a.] International Centre for Quantum and Molecular Structures and the School of Physics, Shanghai University, Shanghai 200444, China.
[b.] School of Mathematical and Physical Sciences, University of Technology Sydney, NSW, 2007, Australia
[c.] School of Mathematics and Physics, The University of Queensland, Brisbane, Queensland 4072, Australia.
[d.] Queensland Micro- and Nanotechnology Centre, Griffith University,
[e.] Nathan, QLD 4111, Australia
[f.] Materials Genome Institute, International Centre for Quantum and Molecular Structures, Shanghai University, Shanghai 200444, China.
* Email: Jeffrey.Reimers@uts.edu.au.

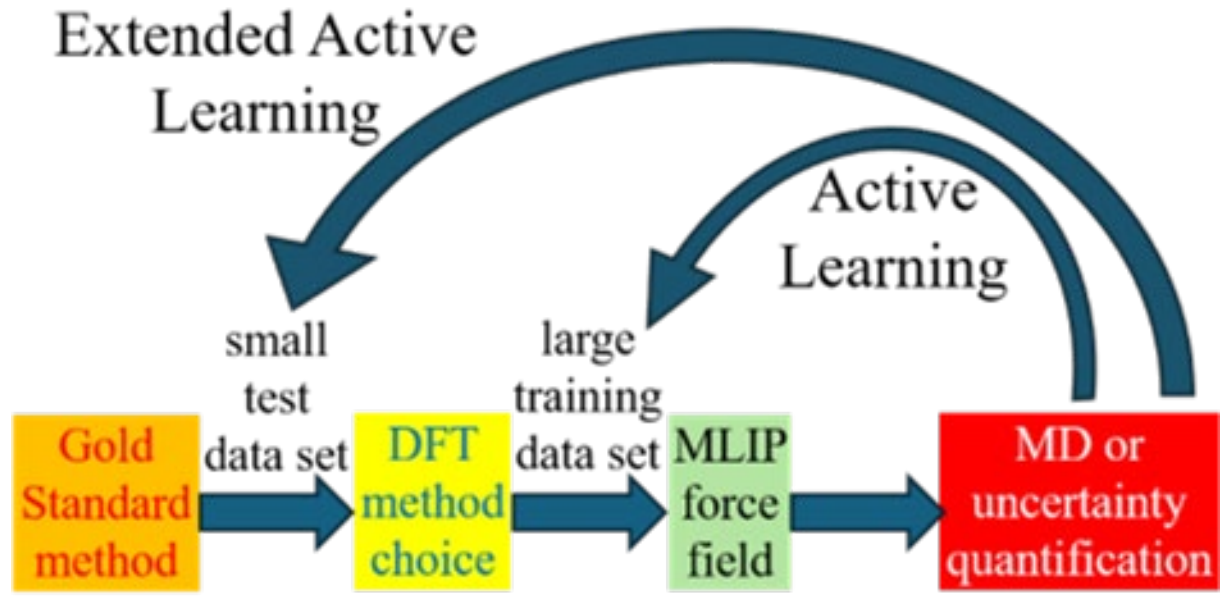

**ABSTRACT:** For the purpose of modelling ferroelectric switching in wurtzite-structured materials, four cost-effective density-functional theory (DFT) methods (PBE, PBEsol, r²SCAN, and r²SCAN-rVV10) are considered and compared to various ab initio approaches based on the random-phase approximation (RPA), including RPA with singles corrections (RPAR+S), as well as second-order Møller-Plesset perturbation theory (MP2). The purpose is to determine whether an ab initio approach could act as a "gold standard" for estimating the reliability of DFT, thus determining an optimal DFT method for use in exhaustive tasks such as the training of machine-learning interatomic potentials (MLIP) for large-scale simulations of materials of arbitrary composition and structure such as $Al_{1-x}Sc_xN$ and $Zn_{1-x}Mg_xO$, using AlN, $Al_{0.5}Sc_{0.5}N$, ZnO, and $Zn_{0.5}Mg_{0.5}O$ as model materials. Applications of active learning (AL) are now common, in which results from MLIP simulations are used to enhance the DFT training data set, but future extended active learning (EAL) methods will also need to systematically assess the DFT methodology against a gold standard. Herein, the variability of the ab initio results is found to exceed that required for a robust gold standard, but the DFT and ab initio approaches appear to converge on RPAR+S and r²SCAN-rVV10 as optimal method choices to initiate EAL. The electron correlation energy is found to be dominated by covalent binding effects associated with the high-electron-density anions involved, but the van der Waals dispersion force is seen to be too significant to ignore in ferroelectric modelling.

## Introduction

Recent demonstrations of ferroelectric switching in doped wurtzite-structured materials have generated much interest for the development of new devices.[1-9] Important materials include Sc-doped AlN, $Al_{1-x}Sc_xN$,[1, 10-13] Mg-doped ZnO, $Zn_{1-x}Mg_xO$,[14-19] and others,[20-23] with research focusing on the distribution of dopants in the material,[15, 24-31] induced changes in crystal structure,[30] coercive field strengths,[12, 32] uncoordinated motions,[23, 33] and the role of domains, grain boundaries and defects in facilitating ferroelectric switching.[1, 8, 9, 11, 13, 20, 21, 34]

Computational methods have aided these endeavours through structural simulations and the modelling of substituent effects, reaction energetics, and polarisation evolution.[11, 12, 14, 20, 21, 23, 34, 35] Traditionally, density-functional theory (DFT) has been the most widely used first-principles approach, owing to its applicability across a broad range of chemical environments. Its application here faces two significant challenges, however: (1) a large number of density functionals are available, each with different advantages and disadvantages, with different functionals often delivering significantly different predictions of structure and energetics relevant to ferroelectric switching;[36] and (2) the dopant concentration $x$ is unconstrained and the dopant distribution may be disordered, requiring large simulation cells and extensive configurational sampling. These challenges compound one another, as DFT approaches that may be more reliable in unusual chemical environments become impractically expensive.

To meet modern needs, computational approaches are now adopting machine learning methods both to identify relevant features[20] and train force fields that can reproduce energetics, forces, and polarisations associated with ferroelectric switching and other phenomena.[37] Universally applicable methodologies based upon DFT have been developed[38] and trained on results obtained using typically the low-level PBE[39] density functional, but its suitability for modelling key aspects of ferroelectric switching remains to be determined. An alternative approach is to train machine-learning interatomic potentials (MLIP) and polarisation models using data specific to the problem of interest.[40-44]

In recent years, advanced model training has employed active learning[45] (AL) techniques that use uncertainty quantification[46, 47] or related selection criteria, often together with molecular dynamics simulations, to provide iterative training[48] approaches in which the training set is progressively refined to better represent newly identified relevant configurations or phenomena.[49] Such approaches, however, still fail to deal with the critical issue that the DFT method used to generate the training data may not be adequate for unusual situations revealed by the active learning. More generally, extended active learning (EAL) techniques involving multiple targets[50] are also required to optimise the choice of DFT method used to train the MLIP (see TOC graphic). This requires a "gold standard" approach to be developed for the purpose of evaluating the suitability of different DFT methods; this would be akin to those already developed for molecular and biochemical applications, including developments by Head-Gordon et al.[51, 52]

The purpose of this work is to seek such a gold standard and hence establish practical and affordable DFT computational methods for MLIP training for the structure and energetics of ferroelectric switching applied to wurtzite-structured materials. Specifically, it focuses on AlN and $Al_{0.5}Sc_{0.5}N$ as model materials, also utilising and extending understanding of the observed structure[53] and computational energetics[36] of ZnO and $Zn_{0.5}Mg_{0.5}O$.

Four cost-effective DFT methods are analysed: a standard generalised-gradient approximation (GGA) functional PBE,[39] its revision intended to improve calculations for solids, PBEsol,[54] the more advanced meta-GGA functional $r^2$SCAN,[55] and the latter augmented with the empirical dispersion correction rVV10[56, 57] to give $r^2$SCAN-rVV10. These methods differ, in particular, in their treatment of the electron correlation and kinetic energies. Hybrid functionals, which improve the treatment of electron exchange, are not used owing to their computational expense in materials-science applications. In particular, treatments for the self-interaction error are not included in our calculations, and the long-range asymptotic potential is not corrected, as it would be partially by using e.g., the PBE0[58] functional, or more fully by using CAM-B3LYP.[59-61]

Consideration of the ubiquitous van der Waals London dispersion force is included only in the $r^2$SCAN-rVV10 calculations. Even though it can control ferroelectric switching in materials containing soft ions,[62, 63] it is mostly neglected in DFT calculations for materials containing hard ions[64, 65] like $O^{2-}$, $Sc^{3+}$, $Mg^{2+}$ and $Al^{3+}$, but the materials of interest herein also contain ions that could be hard or soft, such as $Zn^{2+}$ and $N^{3-}$. For $Zn_{1-x}Mg_xO$, the way that dispersion interactions are treated has been found to influence the way that ferroelectric switching is perceived,[36] with rVV10 being amongst the methods giving the most intuitive results that correspond best with higher-level calculations.

As possible gold-standard benchmarks for evaluating the quality of computationally expedient DFT methods, we consider various ab initio approaches that form part of systematic series leading towards the exact answer. Each individual method in a series may deliver results that are far from the exact limit, however, necessitating rigorous assessment.

We start with the Hartree-Fock (HF) approximation,[66] an approach without self-interaction error but one that ignores critical contributions from electron correlation to covalent bonding and dispersion. This is then improved systematically through use of second-order Møller-Plesset perturbation theory[67, 68] (MP2). Again, there is no self-interaction error, but the contributions of electron correlation to both covalent bonding and dispersion are typically overestimated.[69-72]

A more sophisticated improvement to HF, the random-phase approximation[73-75] (RPA) is also used, empirically adjusted to use orbitals calculated by PBE instead of those from HF. Other calculations are performed using its computationally more expedient approximation RPAR,[76] as well as this augmented by singles corrections, RPAR+S.[77, 78] The singles corrections are formally zero if RPA is started from HF orbitals but otherwise appears as a costly correction, and recent RPA implementations have been developed with an empirical focus that either neglects singles contributions or, in the case of RPAR+S, includes them to first order only. Through the consideration of multiple ab initio approaches, we seek a "gold standard" that could be developed and widely used to evaluate DFT performance. Previous attempts at establishing RPA-based methods in this way have concluded that the method is encouraging but not yet fully adequate,[57, 79] and its appropriateness for ferroelectric switching needs to be established.

We seek test systems for ferroelectric switching that allow ab initio calculations to be used to assess the accuracy of the DFT methods. The high-symmetry materials AlN, $Al_{0.5}Sc_{0.5}N$, ZnO, and $Zn_{0.5}Mg_{0.5}O$ are selected for this purpose. We consider only concerted ferroelectric switching as this simple process is expedient to model. The high-symmetry structures considered herein are useful as they embody both common and unusual chemical scenarios of general relevance when considering local-scale reactions.

Specifically, we consider three main structural types, named **A**, **C**, and **Cv**, as depicted in Fig. 1 for AlN and $Al_{0.5}Sc_{0.5}N$. Structures **A** represent *P*63*mc*-symmetry wurtzite lattices of the materials in which all atoms have tetrahedral coordination. These can be considered as having layers of atoms from which the anions and cations are displaced in opposite directions along the *c* crystallographic axis. The layered structures are named **C** and have no dipole polarisation. Displacements in the two possible directions lead to polarised structures with opposite polarisations, structure **A** and its image **A′**. The structures **C** have *P*63/*mmc* symmetry with trigonal bipyramidal atomic coordination. They may depict transition states for concerted ferroelectric switching or else represent possibly isolatable metastable intermediates protected by transition-state structures later named **B** and **B′**.

The structures **Cv** are variants of **C** in which the lattice vectors are constrained to have the values of the wurtzite structures **A**. They are hypothetical structures relevant to concerted ferroelectric switching in the materials but nevertheless act to mimic unusual situations that may be found during ferroelectric switching associated with defects, grain boundaries, etc. Significantly, in **Cv** the atoms have trigonal planar coordination, paralleling that found in graphene and h-BN, which is common in van der Waals heterostructures. Involvement of such heterostructures in ferroelectric switching therefore brings into prominence the changes that occur in the dispersion interaction during ferroelectric switching.

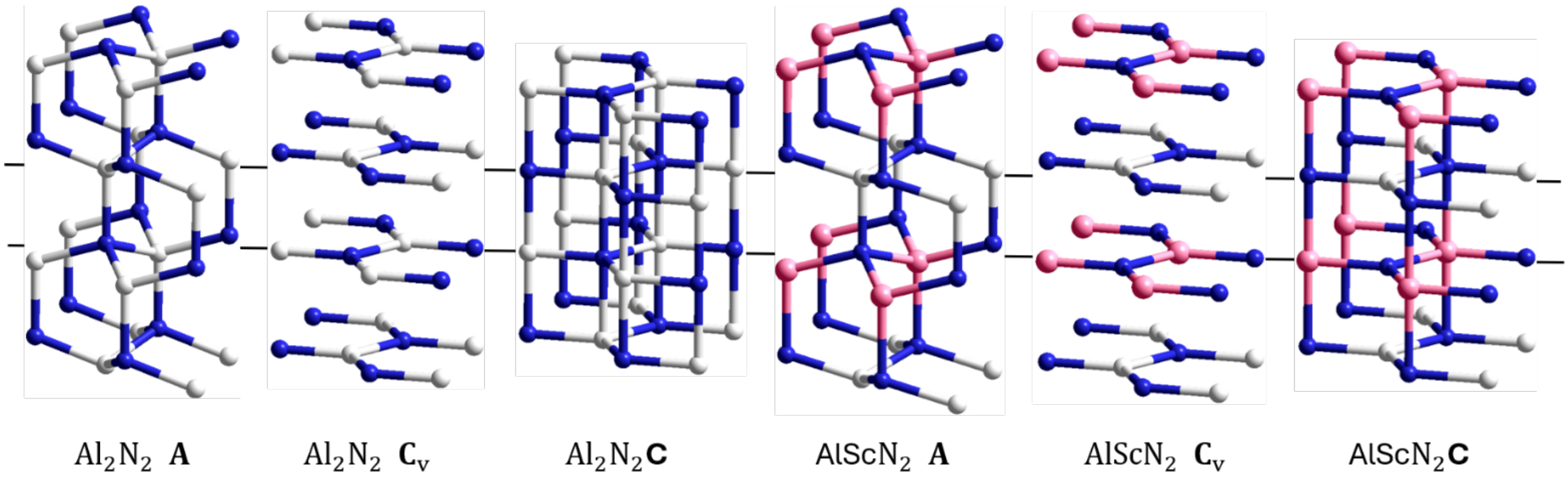

**Fig. 1** Replicas 2 × 2 × 2 of 4-atom unit cells of AlN and $Al_{0.5}Sc_{0.5}N$ in their wurtzite structures **A** (tetrahedral coordination, *P*63*mc*), their associated hexagonal-planar structures **C** (trigonal bipyramidal coordination, *P*63/*mmc*), and their hypothetical hexagonal-planar structures constrained at the lattice vectors of **A**, **Cv** (trigonal planar coordination, *P*63/*mmc*); blue- N, grey- Al, pink- Sc.

## Methods

All calculations were performed using VASP-6.4.3.[80, 81] Most calculations pertained to AlN or $Al_{0.5}Sc_{0.5}N$. For these, "GW" PAW pseudopotentials were used[82] that include 3 valence electrons for Al, 11 for Sc, and 5 for N. In the DFT calculations, the PBE,[39] PBEsol,[54] and r²SCAN[55] density functionals were used, with the rVV10[56, 57] dispersion correction additionally applied to r²SCAN, r²SCAN-rVV10. Calculations were also performed using the HF,[66] MP2,[67] RPA,[73-75] RPAR,[78] and RPAR+S[74, 75] methods at the PBEsol optimised structures. The RPA and RPAR calculations were performed by the recommended procedure that uses PBE orbitals as their input, but test calculations were also performed using rSCAN[83] and r²SCAN orbitals. All calculations were performed using spin-restricted wavefunctions; test calculations using HF and MP2 for $Al_{0.5}Sc_{0.5}N$ using spin-unrestricted wavefunctions yielded equivalent results.

For AlN and $Al_{0.5}Sc_{0.5}N$, the plane-wave energy cutoff was set to 600 eV for the DFT calculations, 520 eV for the MP2 calculations, and 700 eV for the RPA and RPAR calculations. A Monkhorst-Pack 6 × 6 × 4 *k*-point grid[84] was used for most of the Brillouin-zone integrations, with test calculations using denser grids indicating that convergence is generally expected at this level. Alternatively, the MP2 calculations were performed using a 3 × 3 × 2 *k*-point grid and extrapolated to 6 × 6 × 4, as has been demonstrated to be a reliable procedure.[36]

To minimise wrap-around errors, the density of the integration grid was doubled from that required to achieve the stated energy cutoffs. Reciprocal space was used to evaluate the projection operators. The energy convergence criterion was set to at most $10^{-7}$ eV, and structures were optimised to a maximum force of $10^{-4}$ eV/Å. A Gaussian smearing of 0.02 eV was used to weight the occupied orbitals. All calculations and reported energetics pertain to unit cells containing four atoms (see Fig. 1), and crystal structures were optimised individually using the four DFT methods. To maximise computational reproducibility, the optimisation procedure involved repeating each optimisation at least eight times to ensure that the plane-wave basis set used in each individual optimisation corresponded to that appropriate to the final structure.

For ZnO and $Zn_{0.5}Mg_{0.5}O$, all calculations matched conditions used previously for analogous calculations.[36] Method differences to those applied to AlN and $Al_{0.5}Sc_{0.5}N$ include use of a 520 eV for the RPA calculations examining the starting-orbital choice,[36] and the use of PBE0 structures instead of PBEsol.

A specifically designed reaction coordinate *RC* was used to construct the PES, enhancing computational reproducibility. This is given by the ratio of the out-of-plane displacement of the Sc-N bond (or one of the Al-N bonds for AlN) of an arbitrary structure to that for **A**. As all other variables are optimised for each structure, *RC* mimics the phonon mode corresponding to ferroelectric switching for small displacements away from **C**. At larger displacements, *RC* typically differs significantly from that phonon mode as it captures the large effects of anharmonicity that contribute to the reaction. Values of *RC* = 0 correspond to **C**, -1 to **A′**, and +1 to **A**. For AlN, optimisation always led to structures in which the two Al atoms and the two N atoms in the 4-atom unit cell were symmetrically related.

**Table 1.** Calculated properties relevant for ferroelectric switching for 4-atom unit cells of AlN and $Al_{0.5}Sc_{0.5}N$.

| material | method | Unit-cell parameters[a] (Å) | | | | Energy differences (meV) | | | | Phonon[b] |
|---|---|---|---|---|---|---|---|---|---|---|
| | | $a_A$ | $c_A$ | $a_C$ | $c_C$ | $E_C - E_A$ | $E_B - E_C$ | $E_B - E_A$ | $E_{Cv} - E_C$ | (cm$^{-1}$) |
| AlN | PBE | 3.127 | 5.013 | 3.310 | 4.185 | 457 | 6 | 463 | 563 | 199 |
| | PBEsol | 3.113 | 4.980 | 3.300 | 4.138 | 364 | 20 | 384 | 680 | 269 |
| | r²SCAN | 3.104 | 4.974 | 3.290 | 4.131 | 451 | 17 | 468 | 667 | 260 |
| | r²SCAN-rVV10 | 3.101 | 4.968 | 3.287 | 4.126 | 431 | 20 | 451 | 695 | 269 |
| | HF | | | | | 730 | 6 | 736 | 469 | |
| | RPA | | | | | 481 | 10 | 492 | 638 | |
| | RPAR | | | | | 474 | 10 | 484 | 647 | |
| | RPAR+S | | | | | 461 | 9 | 470 | 647 | |
| | MP2 | | | | | 393 | 19 | 412 | 710 | |
| $Al_{0.5}Sc_{0.5}N$ | PBE | 3.336 | 5.191 | 3.537 | 4.375 | 171 | -9[c] | | 779 | 148i |
| | PBEsol | 3.325 | 5.096 | 3.513 | 4.330 | 70 | -3[c] | | 794 | 78i |
| | r²SCAN | 3.317 | 5.161 | 3.519 | 4.344 | 168 | -8[c] | | 876 | 136i |
| | r²SCAN-rVV10 | 3.315 | 5.143 | 3.514 | 4.338 | 143 | -7[c] | | 889 | 123i |
| | HF | | | | | 275 | -10[c] | | 788 | |
| | RPA | | | | | 73 | 2[c] | | 842 | |
| | RPAR | | | | | 78 | -5[c] | | 837 | |
| | RPAR+S | | | | | 123 | -9[c] | | 874 | |
| | MP2 | | | | | -51 | 55 | | 889 | |

a: Observed values for AlN:[85] $a_A$ = 3.112 Å, $c_A$ = 4.979 Å; b: The frequency for **C** of the mode responsible for concerted ferroelectric switching. c: Evaluated at *RC* = 1/4.

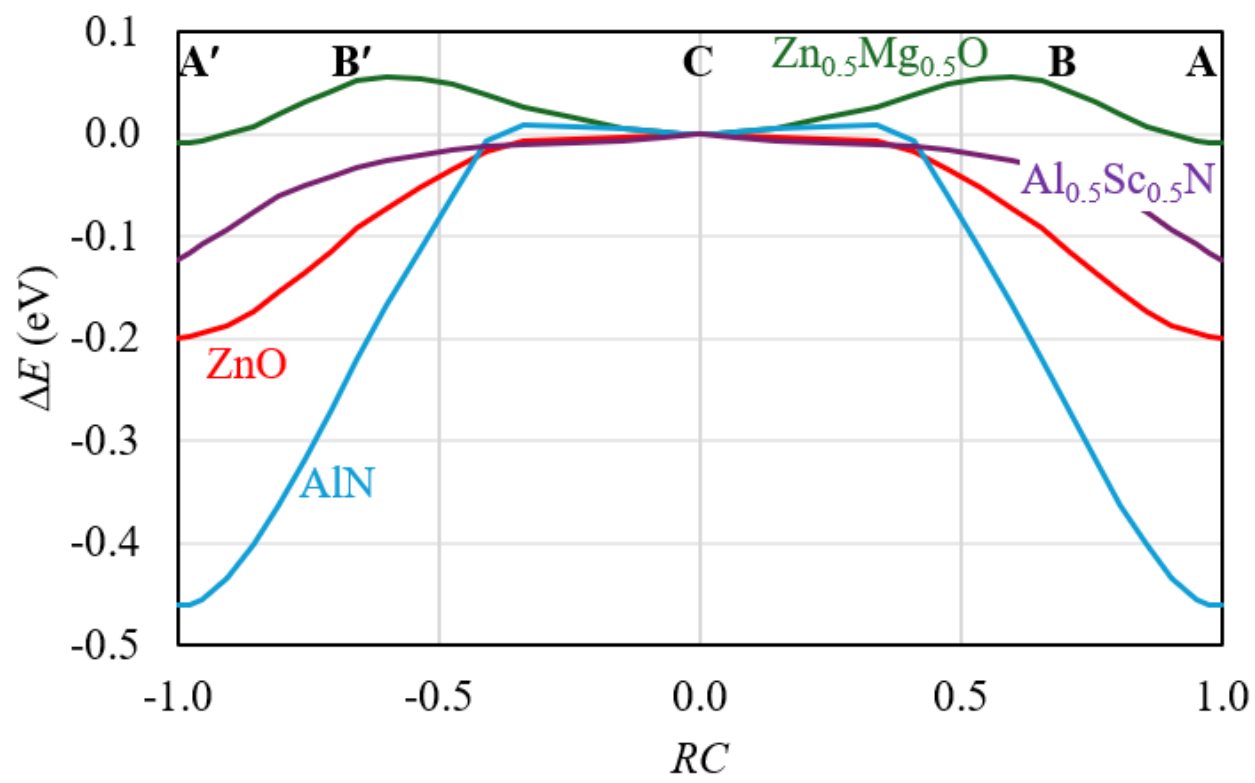

**Fig. 2**. RPAR+S potential-energy surfaces per 4-atom cell for concerted ferroelectric switching in AlN, $Al_{0.5}Sc_{0.5}N$, ZnO, and $Zn_{0.5}Mg_{0.5}O$.

## Results

Extensive computational results are provided in Electronic Supporting Information (ESI), listing up to 48 input or output properties extracted directly from 140 VASP calculations for both AlN and $Al_{0.5}Sc_{0.5}N$; similar results are also provided for ZnO and $Zn_{0.5}Mg_{0.5}O$, either evaluated in this work or previously,[36] as indicated therein. All mathematical manipulations of this data, leading to the main results tables are figures, are included in the ESI, along with additional tabulated and plotted results such as all the optimised geometrical variables plotted as a function of the reaction coordinate *RC*, and various calculations of the materials' band gaps.

Figure 2 highlights the different behaviours predicted by RPAR+S for concerted ferroelectric switching in AlN, $Al_{0.5}Sc_{0.5}N$, ZnO, and $Zn_{0.5}Mg_{0.5}O$, with reaction energy differences and related PBEsol phonon frequencies listed in Table 1 for AlN and $Al_{0.5}Sc_{0.5}N$, (see ESI for ZnO and $Zn_{0.5}Mg_{0.5}O$).

For AlN, the structure C presents as a shallow intermediate phase hexagonal that could possibly be stabilised during ferroelectric switching. It is predicted to be of very high energy relative to A, however, making its isolation difficult.

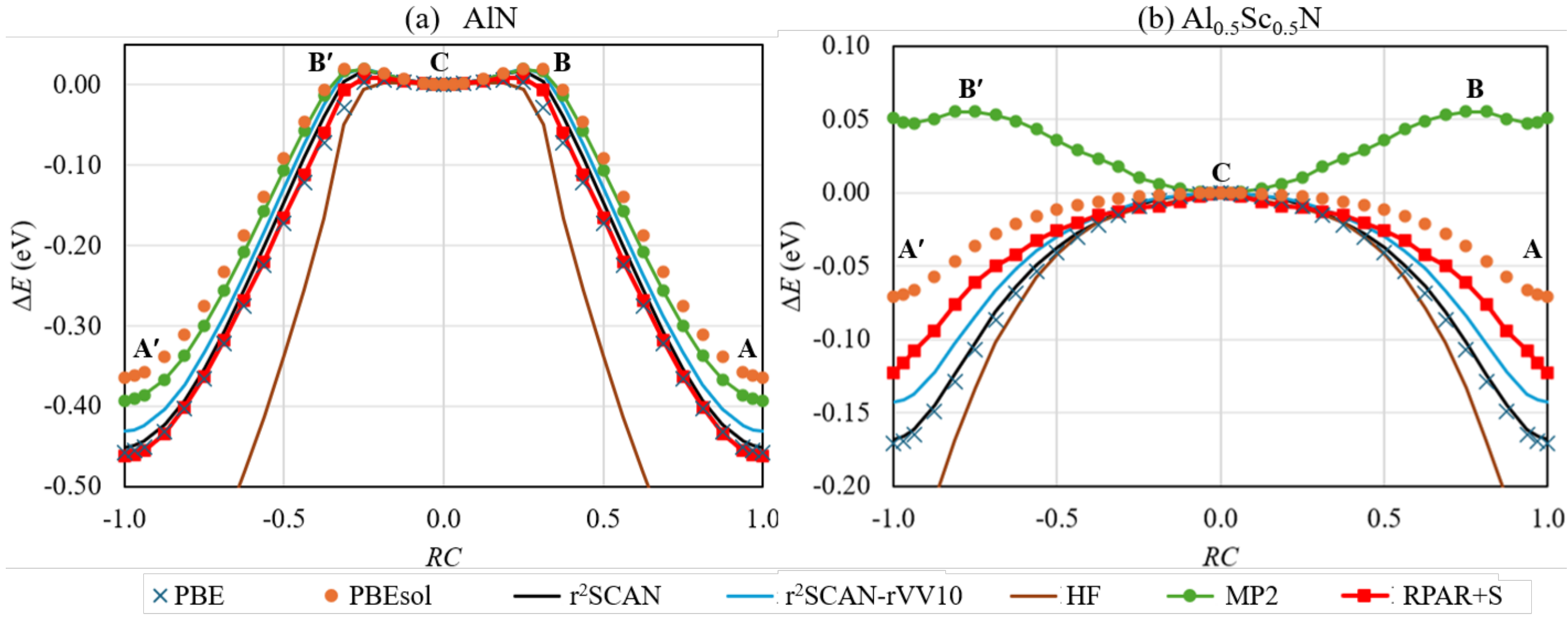

**Fig. 3.** Potential-energy surfaces for concerted ferroelectric switching per 4-atom cell of AlN and $Al_{0.5}Sc_{0.5}N$, as calculated by various DFT and ab initio methods; the HF, MP2, and RPAR+S energies are calculated at PBEsol optimised structures. See ESI for analogous figures for ZnO and $Zn_{0.5}Mg_{0.5}O$.

For $Al_{0.5}Sc_{0.5}N$, the barrier to ferroelectric switching is predicted to be greatly reduced from that in AlN, with **C** as a transition state around which the PES is very flat. To date, neither **C** nor **A** have been identified experimentally, as transformation to a different phase is observed instead for Sc proportions of order 30-37%.[30, 86] Nevertheless, the results indicate that understanding ferroelectric switching in wurtzite-structured materials with similar properties requires highly accurate calculations capable of predicting small energy changes associated with large atomic displacements.

For ZnO, the situation is similar to that for $Al_{0.5}Sc_{0.5}N$ except that structures **A** and **C** have both been observed. Compared to AlN, **A** is relatively less stable compared to **C**, but nevertheless stable enough to prevent the observation of ferroelectric switching. The PES around **C** is predicted to be very flat with **C** appearing as a transition state instead of an intermediate structure. Metastable structures **C** have been observed in nanocrystals,[53, 87] however, and tentatively attributed to be stabilised by surface effects.[36] Again, a central computational challenge is the accurate portrayal of the subtle interactions controlling the PES in the vicinity of **C**.

For $Zn_{0.5}Mg_{0.5}O$, **C** is predicted to be a metastable intermediate, but with an energy only slightly higher than **A**.[30] Ferroelectric switching has been observed in doped ZnO with similar compositions.[14-19] Independent of the detailed mechanism of this process, subtleties such as those depicted in Fig. 2 are likely to remain important.

Figure 3 shows PES calculated by DFT and ab initio methods for AlN and $Al_{0.5}Sc_{0.5}N$, focusing on how these approaches describe the critical chemical subtleties apparent in Fig. 2; key results are also listed in Table 1. Analogous figures and tables for ZnO and $Zn_{0.5}Mg_{0.5}O$ are presented in ESI and elsewhere.[36]

For AlN, the energy difference between the polarised phase **A** and the metastable unpolarised phase **C**, $E_C - E_A$, is very similarly predicted by the PBE, r²SCAN, and RPAR+S methods, with r²SCAN-rVV10 results being close, MP2 results being somewhat different, and PBEsol results being distinctly different. These differences become amplified when considering the shallow barrier stabilising **C**, $E_B - E_C$. The HF calculations fail to include the significant stabilisation of **C** that arises from electron correlation within this high-density structure. As expected, the MP2 correlation energy appears to overestimate this effect relative to methods such as RPAR+S and r²SCAN-rVV10.

For $Al_{0.5}Sc_{0.5}N$, however, the MP2 correlation energy is significantly larger than expected, resulting in the alternative prediction that **C** is more stable than **A**. As neither structure has been observed experimentally owing to an alternative polymorph being more stable, this significant difference cannot currently be assessed. The other methods predict large changes in $E_C - E_A$, indicating that no consistent description of the chemical bonding changes is obtained. These differences have the same absolute magnitude as those predicted for AlN; however, the much smaller magnitude of $E_C - E_A$ for $Al_{0.5}Sc_{0.5}N$ makes small energy changes more significant. The large differences predicted between the RPAR, RPAR+S, and MP2 results lead to considerable uncertainty in the "best estimate" results for the PES: RPAR and PBEsol perceive only a small energy difference $E_C - E_A$ between the phases, PBE and r²SCAN predict a large difference, with RPAR+S and r²SCAN-rVV10 giving intermediate results.

Table 1 also lists DFT-calculated lattice vectors, with ESI listing their deviations from experiment. For AlN, ZnO, and $Zn_{0.5}Mg_{0.5}O$, the mean absolute deviations (MAD) are: 0.011Å for r²SCAN-rVV10, 0.013 Å for PBEsol, 0.015 Å for r²SCAN, and 0.024 Å for PBE; note that, for just ZnO and $Zn_{0.5}Mg_{0.5}O$, reduced errors have previously been predicted by PBE0.[36]

**Table 2.** Analysis of key energetic factors controlling concerted ferroelectric switching (meV).[a]

| material | $E_C - E_A$ | $E_B - E_C$ | $E_B - E_A$ | $E_{Cv} - E_C$ |
|---|---|---|---|---|
| (a) Structural variation of RPAR or MP2 energies[b] | | | | |
| ZnO | 17 | | | 10 |
| $Zn_{0.5}Mg_{0.5}O$ | 57 | | | 49 |
| AlN | 1 | | | 55 |
| $Al_{0.5}Sc_{0.5}N$ | 2 | | | 100 |
| (b) Range of RPA results from orbital-choice variations | | | | |
| ZnO | 32 | 1 | 32 | 9 |
| $Zn_{0.5}Mg_{0.5}O$ | 14 | 4 | 17 | 9 |
| AlN | 15 | 0 | 14 | 8 |
| $Al_{0.5}Sc_{0.5}N$ | 16 | 5 | 11 | 15 |
| (c) Error in the RPAR calculations from full RPA | | | | |
| ZnO | 1 | 0 | 1 | -2 |
| $Zn_{0.5}Mg_{0.5}O$ | -2 | 0 | -2 | 0 |
| AlN | -8 | 0 | -8 | 9 |
| $Al_{0.5}Sc_{0.5}N$ | 5 | -7 | -1 | -6 |
| (d) Effect of the RPAR singles correction | | | | |
| ZnO | -38 | 1 | -37 | 26 |
| $Zn_{0.5}Mg_{0.5}O$ | -31 | 0 | -31 | 4 |
| AlN | -12 | -2 | -14 | -1 |
| $Al_{0.5}Sc_{0.5}N$ | 45 | -5 | 40 | 37 |
| (e) MP2 correlation energy | | | | |
| ZnO | -153 | 9 | -144 | 160 |
| $Zn_{0.5}Mg_{0.5}O$ | -105 | 7 | -98 | 61 |
| AlN | -337 | 13 | -324 | 241 |
| $Al_{0.5}Sc_{0.5}N$ | -326 | 66 | -260 | 100 |
| (f) rVV10 dispersion energy | | | | |
| ZnO | -22 | 2 | -20 | 42 |
| $Zn_{0.5}Mg_{0.5}O$ | -21 | 8 | -13 | 31 |
| AlN | -20 | 2 | -17 | 28 |
| $Al_{0.5}Sc_{0.5}N$ | -25 | 1 | -24 | 13 |

a: Raw data taken from Table 1 and ESI. b: RPA energies for AlN and $Al_{0.5}Sc_{0.5}N$, MP2 energies[36] for ZnO and $Zn_{0.5}Mg_{0.5}O$, see ESI.

Another indicator of the quality of DFT-optimised structures is the energy spread that is obtained when higher-level single-point energy calculations are performed, as listed in Table 2(a). Previously for ZnO and $Zn_{0.5}Mg_{0.5}O$,[36] the lowest-MP2-energy structures were predicted by PBE0 and r$^2$SCAN-rVV10. As a function of the DFT method used to optimise the structure, the calculated ZnO energy ranges were 17 meV for $E_C - E_A$ and 10 meV for $E_{Cv} - E_C$, increasing to 57 meV and 49 meV, respectively, for $Zn_{0.5}Mg_{0.5}O$. Herein, comparing RPA energies obtained at PBE, PBEsol, r$^2$SCAN, and r$^2$SCAN-rVV10 geometries, for AlN we obtain a 1 meV range for $E_C - E_A$ and a 55 meV range for $E_{Cv} - E_C$, changing to 2 meV and 100 meV, respectively, for $Al_{0.5}Sc_{0.5}N$. The ferroelectric-switching pathway therefore appears better represented for AlN and $Al_{0.5}Sc_{0.5}N$, whereas hypothetical van der Waals structures $\mathbf{C_v}$ are poorly represented; for ZnO and $Zn_{0.5}Mg_{0.5}O$, intermediate results were generally obtained. Also, the PBE structures were found to give the lowest RPA energies for AlN and $Al_{0.5}Sc_{0.5}N$, despite PBE giving the poorest agreement with observed lattice parameters. The RPA approach therefore may not be a good choice for the evaluation of optimisation-method quality.

Table 2 lists other metrics obtained from the raw data in Table 1 and ESI that pertain to the reliability of the calculated reaction energetics. Table 2(b) lists the range of RPA energies for ZnO, $Zn_{0.5}Mg_{0.5}O$, AlN, and $Al_{0.5}Sc_{0.5}N$ obtained by varying the DFT orbitals used in the calculations. The orbital variants include PBE orbitals, as recommended, as well as rSCAN and r$^2$SCAN orbitals. The calculated energy ranges vary between 14 meV and 32 meV for $E_C - E_A$, and between 8 meV and 16 meV for $E_{Cv} - E_C$. These ranges are small on the chemical energy scales pertinent to AlN and ZnO but significant for the consideration of $Zn_{0.5}Mg_{0.5}O$ and $Al_{0.5}Sc_{0.5}N$.

Next, Table 2(c) lists the energy difference ranges found when comparing RPA calculations to its low-scaling approximation RPAR. The ranges are typically small, reaching 9 meV at most. This indicates that the RPAR approximation is appropriate for the study of ferroelectricity in these materials.

In contrast, Table 2(d) lists the ranges observed in the singles correction to RPAR. This can be sizable and can vary in sign, ranging from -38 meV to +45 meV. It is therefore essential to determine the appropriateness of the singles correction when applying RPA methods to investigate ferroelectric switching.

As a distinctly different alternative to RPA-based ab initio approaches, MP2 offers insight into calculation reliability. Naively, the MP2 correlation energy is expected to overestimate the contributions of electron correlation to both covalent bonding and van der Waals interactions. The covalent contribution increases with electron density and hence electron correlation favours structure **C** more strongly, followed by **B**, and then **A** and $\mathbf{C_v}$, as is evident from Table 2(e). As AlN is most likely a harder material than ZnO, it would be expected to experience weaker van der Waals interactions, yet the MP2 correlation energy shows the opposite trend, e.g., -337 meV for AlN compared to just -153 meV for ZnO; this indicates that the electron-correlation contribution to $E_C - E_A$ is dominated instead by the effect of the high anionic electron densities.

Table 3. Comparison of DFT energetics to RPAR+S and RPAR (meV).[a]

| material | PBE | PBEsol | $r^2$SCAN | r2SCAN-rVV10 |
|---|---|---|---|---|
| (a) MAD difference DFT to RPAR+S | | | | |
| ZnO | 71 | 30 | 31 | 10 |
| $Zn_{0.5}Mg_{0.5}O^c$ | 66 | 18 | 24 | 9 |
| AlN | 30 | 57 | 10 | 27 |
| $Al_{0.5}Sc_{0.5}N$ | 47 | 25 | 22 | 6 |
| ALL | 52 | 40 | 22 | 14 |
| (g) MAD difference DFT to RPAR | | | | |
| ZnO | 46 | 7 | 8 | 16 |
| $Zn_{0.5}Mg_{0.5}O^c$ | 50 | 16 | 14 | 11 |
| AlN | 26 | 63 | 16 | 33 |
| $Al_{0.5}Sc_{0.5}N$ | 61 | 15 | 35 | 23 |
| ALL | 46 | 28 | 25 | 27 |

a: Raw data taken from Table 1 and/or ESI.

Finally, Table 2(f) lists the rVV10 van der Waals energies calculated for the four materials. The results appear to be similar across the materials, supporting the conclusions drawn about the low significance of the van der Waals electron correlation compared to that associated with covalent bonding. Nevertheless, the dispersion effects range up to 42 meV and are large enough to influence the description of ferroelectric switching in low-barrier materials like $Zn_{0.5}Mg_{0.5}O$ and $Al_{0.5}Sc_{0.5}N$.

Table 3 considers the MAD energy differences for each DFT method and material, as well as for all materials combined, with respect to RPAR+S and RPAR calculations. The largest differences are found mostly for the PBE calculations, ranging from 46 meV to 71 meV, with the exception being the PBEsol calculations for AlN. In contrast, the $r^2$SCAN and $r^2$SCAN-rVV10 approaches yield differences of at most 35 meV and 33 meV, respectively. Considering all data, the smallest MAD is 14 meV for the difference between $r^2$SCAN-rVV10 and RPAR+S. These methods therefore present the greatest convergence of DFT and ab initio approaches.

## Conclusions

No clear gold standard is identified that could be used to gauge DFT performance in an EAL application to ferroelectric switching in $Al_{1-x}Sc_xN$ and $Zn_{1-x}Mg_xO$ materials. This is owing to intrinsic uncertainties in the specification of RPA methodologies, including the choice of orbitals and the appropriateness of the RPA singles correction. Moreover, the conceptually very different MP2 approach, which gave results that strongly supported the RPA results for $Zn_{1-x}Mg_xO$,[36] yields qualitatively different results for $Al_{0.5}Sc_{0.5}N$.

Among the four DFT methods considered, it is clear that both PBE and PBEsol can sometimes deliver results that are broadly inconsistent with the RPA predictions. The robustness of either approach for describing ferroelectric switching is therefore questioned. The most robust approach appears to be $r^2$SCAN-rVV10, which yields results in generally good agreement with RPAR+S.

As a possible rationalisation of this result, it is noted that $r^2$SCAN-rVV10 is the only DFT method considered that includes the van der Waals dispersion force. Also, the singles correction to RPA, considered only in the RPAR+S calculations, directly modulates the treatment of single excitations through a perturbative correction. It is the singles excitations that dominate the polarizability of materials and hence the perceived dispersion force, even at moderate (*i.e.* a few-Å) distances relevant to ferroelectric materials. In our calculations on materials dominated by hard ions, the effect of the dispersion force appears to be small overall, yet it can significantly influence the subtle energetic balance that can control ferroelectric switching. It is therefore recommended that a realistic treatment of the dispersion force be included in all calculations of ferroelectric switching.

Finally, the calculations carried out in this work reveal that true gold-standard benchmarks are urgently required for ferroelectric systems, to provide insights into the quality of (fast) DFT calculations that can provide training data for MLIP and RPA-based calculations that can serve as a faster and more scalable "silver standard" benchmarks. An approach could be based on the recent advances by Lee and Head-Gordon in advancing MP2 to better treat bond-breakage,[88] but this would most likely deliver only another silver standard. The most promising contenders for gold standards are quantum Monte-Carlo[89, 90] (see Ref.[90] and related articles for recent advances) simulations or coupled cluster theory calculations for solids.[91, 92] The extreme sensitivity to changes in electronic structure method revealed in the present work demonstrates that care must be taken to minimize systematic and stochastic errors for predictive accuracy. Work along these lines should be pursued.

## Conflicts of interest

There are no conflicts to declare.

## Data availability

Excel spreadsheets in ESI provide extensive summaries of the inputs and outputs from the VASP calculations, and all mathematical manipulation of the raw results. One workbook is provided for each material considered and one overall summary workbook.

## Acknowledgements

We thank the Australian Research Council for funding this work under the Centre of Excellence in Quantum Biotechnology (QUBIC) grant CE230100021. We thank the National Computational Infrastructure (Australia) and the University of Technology Sydney for providing computational resources through grant d63.

## Notes and references

1. S. Sandeep, R. M. R. Pinto, J. Rudresh, V. Gund, K. K. Nagaraja and K. B. Vinayakumar, *Critical Reviews in Solid State and Materials Sciences*, 2025, **50**, 161-188.
2. M. Belmoubarik, M. Al-Mahdawi, G. Machado, T. Nozaki, C. Coelho, M. Sahashi and W. K. Peng, *Journal of Materials Science: Materials in Electronics*, 2024, **35** 460.
3. C. H. Skidmore, R. J. Spurling, J. Hayden, S. M. Baksa, D. Behrendt, D. Goodling, J. L. Nordlander, A. Suceava, J. Casamento, B. Akkopru-Akgun, S. Calderon, I. Dabo, V. Gopalan, K. P. Kelley, A. M. Rappe, S. Trolier-McKinstry, E. C. Dickey and J. P. Maria, *Nature*, 2025, **637**, 574-579.
4. W. Wang, P. M. Mayrhofer, X. He, M. Gillinger, Z. Ye, X. Wang, A. Bittner, U. Schmid and J. K. Luo, *Appl. Phys. Lett.*, 2014, **105**, 133502.
5. C. Fei, X. Liu, B. Zhu, D. Li, X. Yang, Y. Yang and Q. Zhou, *Nano Energy*, 2018, **51**, 146-161.
6. J. Su, F. Niekiel, S. Fichtner, L. Thormaehlen, C. Kirchhof, D. Meyners, E. Quandt, B. Wagner and F. Lofink, *Appl. Phys. Lett.*, 2020, **117**, 132903.
7. R. M. R. Pinto, V. Gund, R. A. Dias, K. K. Nagaraja and K. B. Vinayakumar, *Journal of Microelectromechanical Systems*, 2022, **31**, 500-523.
8. Y. Zheng, R. Bai, T. Xin, X. Zhao, Y. Cheng, Y.-N. Wu, Y. Wei, B. Ge, H. Tian, S. Chen, Q. Liu, C. Duan and M. Liu, *Science*, 2026, **393**, 85-89.
9. R. Wang, F. Zhu, H. Qian, J. Zhou, W. Sun, S. Zheng, J. Chen, B. Li, Y. Liu, P. Zhou, Y. Hao and G. Han, *Science*, 2026, **393**, 1134-1138.
10. S. Fichtner, N. Wolff, F. Lofink, L. Kienle and B. Wagner, *J. Appl. Phys.*, 2019, **125**, 114103.
11. C. Ke and S. Liu, *arXiv*, 2026, 2604.25343v25341
12. X. Zheng, C. Paillard, D. Wang, P. Chen, H. J. Zhao, Y. Xie and L. Bellaiche, *Phys. Rev. Lett.*, 2026, **136**, 206102.
13. J. Huang, J. Li, X. Guo, T. Wen, D. J. Srolovitz, Z. Chen, Z. Chen and S. Liu, *Phys. Rev. Lett.*, 2026, **136**, 026801.
14. J. Huang, Y. Hu and S. Liu, *Phys. Rev. B*, 2022, **106**, 144106.
15. K. Ferri, S. Bachu, W. Zhu, M. Imperatore, J. Hayden, N. Alem, N. Giebink, S. Trolier-McKinstry and J.-P. Maria, *J. Appl. Phys.*, 2021, **130**, 044101.
16. M. Belmoubarik and A. El Moutaouakil, *Journal of Alloys and Compounds*, 2023, **941** 168960.
17. B. Sharma, R. Gupta, A. Chowdhuri and M. Tomar, *Materials Chemistry and Physics*, 2024, **319** 129375.
18. H. Zhang, A. Alanthattil, R. F. Webster, D. Zhang, M. B. Ghasemian, R. B. Venkataramana, J. Seidel and P. Sharma, *ACS Nano*, 2023, **17**, 17148-17157.
19. R. J. Spurling, D. Goodling, E. Günay, S. S. I. Almishal, E. C. Dickey and J.-P. Maria, *Physical Review Materials*, 2025, **9**, 024405.
20. D. Behrendt, S. Banerjee, J. Zhang and A. M. Rappe, *Physical Review Materials*, 2024, **8**, 055406.
21. B. Dryzhakov, et al., *Adv. Mater.*, 2026, **38**, e20258.
22. J. Su, Z. Xiao, X. Chen, Y. Huang, Z. Lin, J. Chang, J. Zhang and Y. Hao, *npj Computational Materials*, 2025, **11**, 30.
23. S. Calderon, J. Hayden, S. M. Baksa, W. Tzou, S. Trolier-McKinstry, I. Dabo, J.-P. Maria and E. C. Dickey, *Science*, 2023, **380**, 1034-1038.
24. P. Bhattacharya, R. R. Das and R. S. Katiyar, *Appl. Phys. Lett.*, 2003, **83**, 2010-2012.
25. H. Tanaka, S. Fujita and S. Fujita, *Appl. Phys. Lett.*, 2005, **86**, 192911.
26. X. Kang, S. Shetty, L. Garten, J. F. Ihlefeld, S. Trolier-McKinstry and J. P. Maria, *Appl. Phys. Lett.*, 2017, **110**, 042903.
27. C. Xu, X. Pan, H. He and Z. Ye, *Journal of Luminescence*, 2020, **226**, 117456.
28. M. Fujita, M. Sasajima, Y. Deesirapipat and Y. Horikoshi, *J. Cryst. Growth*, 2005, **278**, 293-298.
29. J. Jia, D. Kishi, N. Bai, T. Okajima, F. Lesari and T. Yanagitani, *Phys. Rev. B*, 2024, **109** 134101.
30. X. Zhang, E. A. Stach, W. J. Meng and A. C. Meng, *Nanoscale Horizons*, 2023, **8**, 674-684.
31. B. Bhattarai, X. Zhang, W. Xu, Y. Gu, W. J. Meng and A. C. Meng, *Mater. Horizons*, 2024, **11**, 5402-5408.
32. K. D. Kim, S. K. Ryoo, M. K. Yeom, S. H. Lee, W. Choi, Y. Kim, J.-H. Choi, T. Xin, Y. Cheng and C. S. Hwang, *Nature Communications*, 2025, **16**, 7425.
33. C.-W. Lee, K. Yazawa, A. Zakutayev, G. L. Brennecka and P. Gorai, *Science Advances*, 2024, **10**, eadl0848.
34. E. A. Eliseev, A. N. Morozovska, J.-P. Maria, L.-Q. Chen and V. Gopalan, *Physical Review X*, 2025, **15**, 021058.
35. Y. Kang, W. Zhao, Y. Tong, X. Wang, J. Wang, X. Wang, D. Zhou and M. Yao, *J. Phys. Chem. C*, 2026, **130**, 1401-1411.
36. L. Zhang, M. Li, N. Mehta, C. Verdi, W. Ren and J. R. Reimers, *J. Phys. Chem. C*, 2026, **130**, 7706-7719.
37. I. Batatia, W. J. Baldwin, D. Kuryla, J. Hart, E. Kasoar, A. M. Elena, H. Moore, M. J. Gawkowski, B. X. Shi, V. Kapil, P. Kourtis, I.-B. Magdău and G. Csányi, *arXiv*, 2026, 2602.19411.
38. T. Su, S. Hu, Y. Wu, R. Oyang, X. Wang, M. Li, J. Reimers and T.-Y. Zhang, *arXiv*, 2026, 2603.18650v18651
39. J. P. Perdew, K. Burke and M. Ernzerhof, *Phys. Rev. Lett.*, 1996, **77**, 3865-3868.
40. T. B. Blank, S. D. Brown, A. W. Calhoun and D. J. Doren, *The Journal of Chemical Physics*, 1995, **103**, 4129-4137.
41. J. Behler and M. Parrinello, *Phys. Rev. Lett.*, 2007, **98**, 146401.
42. E. Kocer, T. W. Ko and J. Behler, *Annual Review of Physical Chemistry*, 2022, **73**, 163-186.
43. I. Batatia, et al., *The Journal of Chemical Physics*, 2025, **163**.
44. H. Yang, et al., *arXiv*, 2024, 2405.04967.
45. H. Xu, T. Cui, C. Tang, J. Ma, D. Zhou, Y. Li, X. Gao, X. Gong, W. Ouyang, S. Zhang and M. Su, *Nature Communications*, 2025, **17**, 937.
46. J. Zhang, J. Yin and R. Wang, *Mathematical Problems in Engineering*, 2020, **2020**, 6068203.
47. R. C. Smith, ed., *Uncertainty Quantification: Theory, Implementation, and Applications*, Society for Industrial and Applied Mathematics, Philipadelphia USA, 2024.
48. K. Kang, T. A. R. Purcell, C. Carbogno and M. Scheffler, *Physical Review Materials*, 2025, **9**, 063801.
49. D. Teney, Y. Lin, S. J. Oh and E. Abbasnejad, presented in part at the Advances in Neural Information Processing Systems 36, 2023 71703-71722.
50. S. J. Huang, Y. Li and Y. P. Tang, *IEEE Transactions on Pattern Analysis and Machine Intelligence*, 2026, **48**, 140-154.
51. L. W. Bertels, J. Lee and M. Head-Gordon, *Journal of Chemical Theory and Computation*, 2021, **17**, 742-755.
52. J. Liang and M. Head-Gordon, *Journal of Chemical Theory and Computation*, 2025, **21**, 12601-12621.
53. M. Li, L. Zhang, W. Ren and J. R. Reimers, *Acta Crystallographica Section B*, 2026, **82**, 310-315.
54. J. P. Perdew, A. Ruzsinszky, G. I. Csonka, O. A. Vydrov, G. E. Scuseria, L. A. Constantin, X. Zhou and K. Burke, *Phys. Rev. Lett.*, 2008, **100**, 136406.
55. J. W. Furness, A. D. Kaplan, J. Ning, J. P. Perdew and J. Sun, *J. Phys. Chem. Lett.*, 2020, **11**, 8208-8215.
56. R. Sabatini, T. Gorni and S. de Gironcoli, *Phys. Rev. B*, 2013, **87**, 041108.
57. M. Kothakonda, A. D. Kaplan, E. B. Isaacs, C. J. Bartel, J. W. Furness, J. Ning, C. Wolverton, J. P. Perdew and J. Sun, *ACS Materials Au*, 2023, **3**, 102-111.
58. J. P. Perdew, M. Ernzerhof and K. Burke, *J. Chem. Phys.*, 1996, **105**, 9982-9985.
59. T. Yanai, D. P. Tew and N. C. Handy, *Chem. Phys. Lett.*, 2004, **393**, 51-57.

60 M. Li, R. Kobayashi, R. D. Amos, M. J. Ford and J. R. Reimers, *Chem. Sci.*, 2022, **13**, 1492-1503.
61 M. Li, J. R. Reimers, M. J. Ford, R. Kobayashi and R. D. Amos, *J. Comput. Chem.*, 2021, **42**, 1486-1497.
62 S. A. Tawfik, J. R. Reimers, C. Stampfl and M. J. Ford, *J. Phys. Chem. C*, 2018, **122**, 22675-22687.
63 J. R. Reimers, S. A. Tawfik and M. J. Ford, *Chemical Science*, 2018, **9**, 7620-7627.
64 R. G. Pearson, *Journal of the American Chemical Society*, 1963, **85**, 3533-3539.
65 R. G. Pearson, *Accounts of Chemical Research*, 1993, **26**, 250-255.
66 V. Fock, *Zeitschrift für Physik*, 1930, **62**, 795-805.
67 C. Møller and M. S. Plesset, *Phys. Rev. A*, 1934, **46**, 618.
68 M. Marsman, A. Grüneis, J. Paier and G. Kresse, *J. Chem. Phys.*, 2009, **130**, 184103.
69 T. Schwabe and S. Grimme, *Accounts of Chemical Research*, 2008, **41**, 569-579.
70 J. Shee, M. Loipersberger, A. Rettig, J. Lee and M. Head-Gordon, *The Journal of Physical Chemistry Letters*, 2021, **12**, 12084-12097.
71 N. Mehta, M. Casanova-Páez and L. Goerigk, *Physical Chemistry Chemical Physics*, 2018, **20**, 23175-23194.
72 R. Sedlak, T. Janowski, M. Pitoňák, J. Řezáč, P. Pulay and P. Hobza, *Journal of Chemical Theory and Computation*, 2013, **9**, 3364-3374.
73 D. Bohm and D. Pines, *Phys. Rev.*, 1953, **92**, 609-625.
74 J. Harl and G. Kresse, *Phys. Rev. Lett.*, 2009, **103**, 056401.
75 J. Harl, L. Schimka and G. Kresse, *Phys. Rev. B*, 2010, **81**, 115126.
76 M. Kaltak, J. Klimeš and G. Kresse, *Phys. Rev. B*, 2014, **90**, 054115.
77 X. Ren, P. Rinke, G. E. Scuseria and M. Scheffler, *Phys. Rev. B*, 2013, **88**, 035120.
78 J. Klimeš, M. Kaltak, E. Maggio and G. Kresse, *J. Chem. Phys.*, 2015, **143**, 102816.
79 M. Kim, W. J. Kim, T. Gould, E. K. Lee, S. Lebègue and H. Kim, *J. Am. Chem. Soc.*, 2020, **142**, 2346-2354.
80 G. Kresse and J. Hafner, *Phys. Rev. B*, 1993, **47**, 558-561.
81 G. Kresse and J. Furthmüller, *Comput. Mat. Sci.*, 1996, **6**, 15-50.
82 G. Kresse and D. Joubert, *Phys. Rev. B*, 1999, **59**, 1758.
83 A. P. Bartók and J. R. Yates, *J. Chem. Phys.*, 2019, **150**, 161101.
84 H. J. Monkhorst and J. D. Pack, *Phys. Rev. B*, 1976, **13**, 5188.
85 I. Vurgaftman and J. R. Meyer, *J. Appl. Phys.*, 2003, **94**, 3675-3696.
86 S. Satoh, K. Ohtaka, T. Shimatsu and S. Tanaka, *J. Appl. Phys.*, 2022, **132**.
87 C. Lizandara Pueyo, S. Siroky, S. Landsmann, M. W. E. van den Berg, M. R. Wagner, J. S. Reparaz, A. Hoffmann and S. Polarz, *Chem. Mat.*, 2010, **22**, 4263-4270.
88 J. Lee and M. Head-Gordon, *Journal of Chemical Theory and Computation*, 2018, **14**, 5203-5219.
89 W. M. C. Foulkes, L. Mitas, R. J. Needs and G. Rajagopal, *Reviews of Modern Physics*, 2001, **73**, 33-83.
90 M. A. Morales-Silva, K. D. Jordan, L. Shulenburger and L. K. Wagner, *The Journal of Chemical Physics*, 2021, **154**.
91 A. P. Bartók, J. Kermode, N. Bernstein and G. Csányi, *Physical Review X*, 2018, **8**, 041048.
92 V. A. Neufeld, H.-Z. Ye and T. C. Berkelbach, *The Journal of Physical Chemistry Letters*, 2022, **13**, 7497-7503.